\documentclass[a4paper,superscriptaddress, nofootinbib,showkeys,floatfix,12pt]{revtex4-1}
\usepackage[utf8x]{inputenc}
\usepackage{amsmath}
\usepackage{amsfonts}
\usepackage{amssymb}
\usepackage{amsthm}
\usepackage{mathrsfs}
\usepackage{multirow}
\usepackage{color}
\usepackage[caption=true]{subfig}
\usepackage{array}
\usepackage{subfig}
\usepackage{tikz}
 \usepackage{slashed}
\usepackage{amsmath,amssymb}
\usetikzlibrary{calc}
\usetikzlibrary{decorations}
\usetikzlibrary{decorations.fractals}
\usetikzlibrary{decorations.pathreplacing}
\usetikzlibrary{decorations.pathmorphing}
\usetikzlibrary{decorations.markings}
\usetikzlibrary{positioning,matrix}
\usetikzlibrary{shapes.arrows}
\usetikzlibrary{shapes.symbols}
\usetikzlibrary{shapes.misc}
\usetikzlibrary{shapes.geometric}
\usetikzlibrary{through}
\usetikzlibrary{mindmap,backgrounds}
\usetikzlibrary{fit}
\usetikzlibrary{arrows,positioning}
\usetikzlibrary{external}
\usepackage[colorlinks, urlcolor=black, citecolor=blue, link
color=black]{hyperref} 
\def\dsp{\displaystyle}

\def\nn    {\nonumber}
\def\sss{\scriptscriptstyle}

\newcommand{\bracket}[3]{\langle #1 \lvert #2 \rvert #3 \rangle}
\newcommand{\modulus}[1]{\left| #1 \right|}
\allowdisplaybreaks

\begin{document}
\title{Probing anomalous Higgs couplings in $H \to Z V$ decays}

\author{Tanmoy Modak}
\email{tanmoyy@imsc.res.in}
\author{Rahul Srivastava}
\email{rahuls@imsc.res.in}
\affiliation{The Institute of Mathematical Sciences, Taramani,
Chennai 600113, India}
\date{\today}

\begin{abstract}
We analyze the possibility of probing anomalous  Higgs couplings in the rare decays 
$H \to Z V$, $V$ being a vector quarkonium state.
These rare decays involve both gauge as well as the Yukawa sectors and either of them 
can  potentially be anomalous. 
We show that the branching fractions for $H \to Z V$ decays
in Standard Model are small, making it a sensitive probe for anomalous Higgs couplings
originating from physics beyond Standard Model. 
 Moreover, as both $V$ and $Z$ can decay into pair of charged leptons, 
they provide experimentally clean channels and future LHC runs should observe such decays. 
 We perform a model independent analysis and show how 
angular asymmetries can be used to probe these anomalous Higgs couplings, 
taking further decays of  $V$ and $Z$ to pair of charged leptons into account. 
The angular asymmetries can provide significant information about anomalous Higgs couplings
in both gauge and Yukawa sectors. 
\end{abstract}


\maketitle


\section{Introduction}
\label{sec:intro}


The ATLAS and CMS collaborations at Large Hadron Collider (LHC) have
recently discovered a new bosonic resonance of mass around 125 GeV 
\cite{:2012gk, ATLAS:science,:2012gu,CMS:science, :2012br}. Measuring its coupling 
to different  Standard Model (SM) particles and establishing its nature are going to 
be leading aims of future LHC runs. 
Any deviations from its SM  nature should exhibit in its coupling to 
different particles. Anomalous couplings of Higgs\footnote{Although it is yet to be confirmed 
as SM Higgs, for sake of brevity, in this paper we specify this resonance as Higgs and denote it by $H$.} may come in both gauge and Yukawa
sectors. Establishing the nature of the Higgs will require a precise
measurement of its gauge as well as Yukawa couplings. In future LHC runs the coupling of Higgs to $W, Z$ bosons 
will be measured in different decay modes such as $H \to Z Z^{*}\to 4 \ell$ and will provide us a good understanding of its gauge structure.
However, measuring its coupling to fermions as well as loop induced couplings like $HZ\gamma$ are going to be 
relatively more challenging. Further investigations, both theoretical and experimental, are required to 
find novel ways to explore such couplings. 

The accurate determination of the Higgs Yukawa couplings through direct detection i.e. via $H \to q \bar{q}$
decay modes, are very challenging due to the overwhelming QCD backgrounds. 
Therefore, it is imperative to look for other ways to probe these couplings. In this regard rare decays of Higgs provide an 
excellent alternate probe for measuring them. As we discuss in details in this paper, 
they offer complimentary information about Higgs couplings~\cite{Isidori:2013cla} and can serve as important probe of ``New Physics'' (NP).
Due to their importance, several recent studies \cite{Isidori:2013cla,Gonzalez-Alonso:2014rla,
Buchalla:2013mpa,Gao:2014xlv, Bhattacharya:2014rra,Keung:1983ac,Hagiwara:1993sw,Curtin:2013fra,Bodwin:2014bpa,Korchin:2014kha,
Delaunay:2013pja,Giudice:2008uua,Kagan:2014ila,Chen:2014gka,Beneke:2014sba,Manohar:2000dt,
Isidori:2013cga,Brod:2013cka,Falkowski:2014ffa,Bergstrom:1985hp,Grinstein:2013vsa,Bodwin:2013nua,Modak:2016cdm} 
have been directed towards rare Higgs decays.  

Although the branching ratios are small, rare Higgs decay  rates are enhanced by resonant production 
of $V$ and they could be seen in high luminosity LHC runs or in future colliders. Among rare Higgs decays, 
the decays $H \to Z V$; $V$ being a vector quarkonium e.g. $ J/\psi, \Upsilon$($J^{\rm{PC}} = 1^{--}$) have 
received considerable attention in recent times ~\cite{Gonzalez-Alonso:2014rla,Gao:2014xlv, Bhattacharya:2014rra}. 
As we explicitly show in this work, in SM the branching fractions of these rare decays are very small. However, 
they can be significantly enhanced by new physics contributions, making $H \to Z V$ decays very sensitive probes 
for search of physics beyond SM. Besides, subsequent decays of $Z$ and $V$ into  pair of leptons make them 
experimentally clean channels. Thus, they are important channels to probe anomalous Higgs Yukawa couplings originating from new physics contributions.

Owing to the importance of rare decays of the Higgs, the ATLAS collaboration~\cite{Aad:2015sda} has recently performed an analysis on 
$H \to J/\psi \, \gamma$ and $H \to \Upsilon \gamma$ decay channels and has put limits on the branching fraction of
such decays. Angular analysis of these decays would allow us to infer the nature of $H q \bar{q}$ couplings in a
relatively easier way than the direct $H \to q \bar{q}$ study. However, as we point out in this work, the 
$H \to Z V$ decays allows one to construct several other angular observables due to the subsequent decay
of both $V$ and $Z$ into pair of charged leptons. Thus the studies of $H\to Z V$ decays will be phenomenologically 
richer than $H \to \gamma V$ decays and can give more information about the magnitude as well as sign of the Yukawa couplings of $H$ to heavy quarks.
 
In SM there exist three different channels that contribute to the $H \to Z V$ decay i.e. 
$H \to Z^* Z$ with $Z^* \to V$, $H \to Z \gamma^{*} \to Z V$ and $H \to q \bar{q}\to Z V$. 
The first channel involves Higgs decay to an on-shell $Z$ along with an off-shell $Z^*$. 
The off-shell $Z^*$ then further decays to a $q \bar{q}$ pair which ultimately hadronizes to a vector quarkonium ($J/\psi$ or $\Upsilon$). 
The second channel involves loop induced decay of Higgs to an on-shell $Z$ along with a $\gamma^*$ which 
further decays to $q \bar{q}$, finally hadronizing to give $V$. As we show in this work, although in SM the 
channel $H \to Z \gamma^{*} \to Z V$  is loop suppressed, it can still provide a significant 
contribution depending on the nature of the vector boson $V$.
The third contribution to $H \to Z V$ decays is via $H \to q \bar{q}\to Z V$ channel, which is sensitive to 
the $H$ coupling to quarks coming from the Yukawa sector. Thus $H \to Z V$ decays are not only sensitive 
to Higgs Yukawa couplings but provide an independent probe to the anomalous Higgs gauge 
couplings originating from new physics contributions. As any of the above 
three channels could be anomalous, in this paper we perform a model independent analysis of $H\to Z V$ decay 
without making any assumption on the origin of anomalous couplings. In our work we consider all possible sources that can contribute 
to the $HZV$ vertex for both $\Upsilon$ and $J/\psi$ states.

As mentioned before, both $Z$ and $V$ will further decay to a pair of charged leptons which make $H \to Z V$ 
decays experimentally clean channels to probe. Furthermore, this also allows us to fully reconstruct the four momenta of $H$. 
This facilitates us to construct several angular distributions in terms of different kinematic variables. 
In particular, one can construct two polar angles ($\theta_1$ , $\theta_2$ ) from the pair of 
leptons coming from decay of $Z$ and $V$ repectively along with an azimuthal angle ($\phi$) between
the planes of the two lepton pairs. 

In our work we take a model independent approach and write down the most general $H  Z V$ vertex.
We then derive the angular distributions for $H \to Z V\to 4 \ell$ decays. We show how 
to extract independent angular observables in terms of angular asymmetries from the three angular distributions.
Study of these observables offer unique probe to the CP structure of $HZV$ decays. 
Moreover as the angular observables are functions of $HZV$ couplings therefore any hint of 
NP in the $HZV$ vertex can be extracted via them. These asymmetries have 
been discussed in \cite{Buchalla:2013mpa,Modak:2013sb,Modak:2014zca,Bhattacherjee:2015xra} in the context of 
$H \to Z Z^*\to 4 \ell$ decays, to probe non standard Higgs coupling via angular analysis. They provide powerful tools which 
can probe SM as well as any anomalous contributions to the rare Higgs decays. 
In our work we construct all possible asymmetries and perform a case by case 
analysis discussing relative contributions of different diagrams and the effect 
of anomalous couplings in gauge or Yukawa sector on these asymmetries.

The plan of the paper goes as follows. In section~\ref{sec:direct} we compute the 
SM contributions of the three channels and compare their relative strengths. 
Section~\ref{sec:anghvz} is devoted to formalization of the angular analysis and 
construction of angular asymmetries for $H \to Z V$ with further decays of 
$Z$ and $V$ into pair of charged leptons taken into account. We also discuss how to 
probe different Higgs couplings using these angular asymmetries. 
In Section~\ref{sec:conclusion} we conclude our results.


\section{Standard Model contribution of different channels to $H \to Z V $ decays}
\label{sec:direct}


We start our discussion by first estimating the relative strength of SM contribution of different channels to
the process $H \to Z V$, where $V$ is a vector quarkonium ($J^{PC}=1^{--}$). 
A precise estimation of the SM contribution is needed 
as a precursor to any discussion of new physics contribution in such decays. 
In particular we will focus on $J/\psi(1S)$ 
and $\Upsilon(1S)$ but our analysis is general and 
can be used for any vector quarkonium resonance. 

\begin{figure*}[hbtp!]
\centering
\includegraphics[width=.9\textwidth]{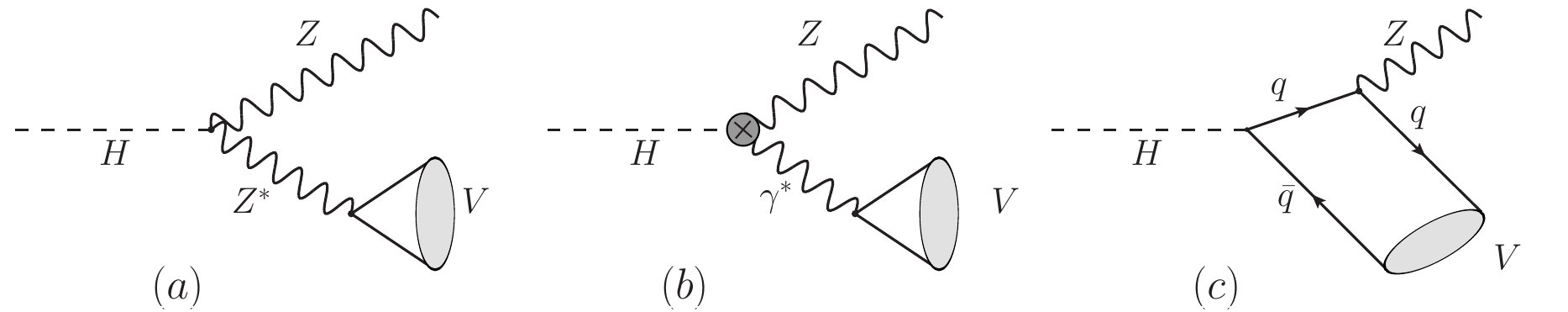}
\caption{Feynman diagrams contributing to $H \to Z V$, $V$ being a vector quarkonium resonance. 
The diagrams originate from three different couplings: (a)  tree level $HZZ$ coupling,
 (b) loop induced $HZ\gamma$ coupling, (c)  $H q \bar{q}$ Yukawa coupling.}
\label{fig1}
\end{figure*}

These decays receive contributions from three different diagrams as shown in Fig.\ref{fig1}. 
To calculate the correct SM contributions one needs to take all the three 
channels as well as the interference terms into account.  Some of these
 contributions have been individually studied in recent works 
\cite{Gao:2014xlv, Bhattacharya:2014rra}. However, the authors of these 
papers have made implicit assumptions regarding the insignificance of certain 
channels and have neglected their contributions. As we show in this section, 
such assumptions are not always justified and may lead to erroneous estimation 
of the branching fraction of these decays.  Thus the primary aim of this section 
is to perform a  combined analysis by correctly including all the channels
and the contributions from the interference terms. Such an analysis is 
still lacking in literature.

The relative strength of the diagrams and their interference terms 
vary depending on the final vector quarkonium resonance. Because of quite different masses of 
$J/\psi$ and $\Upsilon$ resonances, the relative
strengths of these diagrams differ appreciably in the two cases.
We explicitly calculate the individual contributions for $J/\psi (1S)$ and $\Upsilon(1S)$
to demonstrate this fact.

The first diagram Fig.\ref{fig1}(a), originates from tree level $HZZ$ gauge coupling. 
The Lorentz invariant and gauge invariant matrix element can be written as
\begin{align}
\mathcal{M}_1 & = -\mathcal{K}_1\left(a^{ZZ}_1 g_{\mu \nu}+ a^{ZZ}_2\left( q_1\cdot q_2\;g_{\mu \nu} - q_{2\mu}q_{1\nu}\right) 
                            + i a^{ZZ}_3 \,\epsilon_{\mu\nu\rho\sigma} \;  q_2^{\rho}\,q_1^{\sigma}\right)\epsilon_1^{*\mu}\epsilon_2^{*\nu}
\end{align}
where
\begin{align}
 \mathcal{K}_1=\frac{2 g~g^q_V f_V}{v \cos\theta_W}\frac{M_V M_Z^2}{M_Z^2-M_V^2}\label{k1},
\end{align}
with $\theta_W$ as Weinberg angle, $g^q_V=(\frac{1}{4}-\frac{2}{3}\sin^2\theta_W)$ for charm($c$) quark and
 $g^q_V=(-\frac{1}{4}+\frac{1}{3}\sin^2\theta_W)$ for bottom($b$) quark. Also,  $\epsilon_1^{\mu}(q_1)$ and $\epsilon_2^{\nu}(q_2)$ 
are the polarization vectors for $Z$ and $V$ having momenta $q_1$ and $q_2$ respectively. Moreover, 
$f_V$ is defined by the matrix element $\bracket{0}{\bar{q}\gamma^{\mu}q}{V(q_2,\epsilon_2)}=f_V M_V \epsilon_2^{\mu}$.
In the SM at tree level $a^{ZZ}_1=1$ and $a^{ZZ}_2=a^{ZZ}_3=0$. It should be noted that in the above parametrization
$a^{ZZ}_2$ and $a^{ZZ}_3$ have mass dimension $-2$.

Since in SM the $HZ\gamma$ coupling is forbidden at tree level, the second diagram Fig.\ref{fig1}(b),
can only arise via loop processes. One can estimate the contribution of this diagram by writing down an effective Lagrangian 
for the $HZ\gamma$ coupling \cite{Korchin:2013ifa,Bergstrom:1985hp,Hagiwara:1993sw}.
The most general Lorentz invariant and gauge invariant matrix element  for this diagram is given by 
\begin{align}
\mathcal{M}_2 &=-\mathcal{K}_2 \left(a^{Z\gamma}_1 ~\left(q_1.q_2 ~g_{\mu \nu}- q_{1\nu}q_{2\mu}\right)+
                           i a^{Z\gamma}_3\,\epsilon_{\mu\nu\rho\sigma} \; q_2^{\rho}\,q_1^{\sigma}\right)
\epsilon_1^{*\mu}\epsilon_2^{*\nu}
\end{align}
where 
\begin{align}
\mathcal{K}_2 = \frac{g\,\alpha\, Q^f f_V}{2 \pi v} \frac{C_{Z\gamma}}{M_V},
\end{align}
$C_{Z\gamma}$ is the dimensionless effective
coupling constant for the $HZ\gamma$ vertex \cite{Bergstrom:1985hp,Hagiwara:1993sw,Korchin:2013ifa}, $\alpha = \frac{e^2}{4 \pi}$ and 
$Q^f = \frac{2}{3}, \frac{-1}{3}$ for $V = J/\psi, \Upsilon$ respectively. In the SM $a^{Z\gamma}_1=1$ and $a^{Z\gamma}_3=0$.

 Fig.\ref{fig1}(c) comes from $Hq\bar{q}$ Yukawa coupling and the 
corresponding Lorentz and gauge invariant matrix element is given by
\begin{align}
\mathcal{M}_3 &=-\mathcal{K}_3 \left(a^{Zq\bar{q}}_1 ~\left(q_1.q_2 ~g_{\mu \nu}- q_{2\mu} q_{1\nu}\right)+
                                     i a^{Zq\bar{q}}_3\,\epsilon_{\mu\nu\rho\sigma} \; q_2^{\rho}\,q_1^{\sigma}\right)
\epsilon_1^{*\mu}\epsilon_2^{*\nu}
\end{align}
where 
\begin{align}
\mathcal{K}_3 = \frac{4\sqrt{3} g g^q_V \phi_0}{\cos \theta_W\left(M_H^2-M_Z^2-M_V^2\right)}\left(\frac{M_V G_F}{2\sqrt{2}}\right)^{\frac{1}{2}},
\end{align}
and $\phi_0$ is the wave function of the
vector quarkonium resonance evaluated at zero three momentum \cite{Bodwin:2013gca,Bodwin:1994jh,Bodwin:2014bpa}.
In the SM at tree level $a^{Zq\bar{q}}_1=1$ and $a^{Zq\bar{q}}_3=0$.

The total decay width for $H \to Z V$ decays are combinations of all three contributions
 given by
\begin{align}
\Gamma_{total}& =\Gamma_{11}+\Gamma_{22}+\Gamma_{33}+\Gamma_{12}+\Gamma_{13}+\Gamma_{23}.
\end{align}
where $\Gamma_{ii}$ are obtained from $|\mathcal{M}_i|^2$ and $\Gamma_{ij}$
are interference terms between $\mathcal{M}_i$ and $\mathcal{M}_j$ with $i,j=1,2,3$. The individual
contributions for both $ J/\psi(1S) $ and $\Upsilon(1S)$ are listed in Table~\ref{tab}.
\begin{table}[htb]
\caption{Contributions to the branching fraction from the three contributing diagrams and their interferences for $J/\psi(1S)$ 
and $\Upsilon(1S)$ resonances. The total decay width of Higgs is taken to be $4.07$ MeV. We have taken $f_V=0.405(0.680)$ GeV
\cite{Gonzalez-Alonso:2014rla} and $\phi^2_0=0.073(0.512)$ $\rm{GeV}^3$\cite{Bodwin:2013gca}  for $J/\psi(\Upsilon)$.}

\begin{tabular}{c c c}\hline\hline 
$\mathcal{B}r(H \to Z V)$  \hspace{1cm} & $ J/\psi(1S) $                    \hspace{1cm}    & $\Upsilon(1S)$  \\ \hline 
$\mathcal{B}r_{\Gamma_{11}}$\footnote{We define $\mathcal{B}r_{\Gamma_{ij}} = \frac{\Gamma_{ij}}{\Gamma}$ where $\Gamma$ is the total decay width. Note that $\mathcal{B}r_{\Gamma_{ij}}$ is not an observable quantity.}          \hspace{1cm}  &  $1.75 \times 10^{-6}$     \hspace{1cm}   & $1.68 \times 10^{-5}$ \\ 
$\mathcal{B}r_{\Gamma_{22}}$                       \hspace{1cm} & $1.14 \times 10^{-6}$       \hspace{1cm} &  $8.33 \times 10^{-8}$ \\ 
$\mathcal{B}r_{\Gamma_{33}}$                       \hspace{1cm}  & $8.52 \times 10^{-9}$       \hspace{1cm} &  $5.80 \times 10^{-7}$ \\ 
$\mathcal{B}r_{\Gamma_{12}}$                      \hspace{1cm}  & $4.50 \times 10^{-7}$       \hspace{1cm} & $1.10\times 10^{-6}$ \\ 
$\mathcal{B}r_{\Gamma_{13}}$                      \hspace{1cm} & $3.89 \times 10^{-8}$       \hspace{1cm} &  $2.89 \times 10^{-6}$ \\ 
$\mathcal{B}r_{\Gamma_{23}}$                      \hspace{1cm} & $1.97\times 10^{-7}$       \hspace{1cm} & $4.40\times 10^{-7}$ \\ \hline \hline 
\end{tabular}
 \label{tab}
\end{table}

From Table\,\ref{tab} it is clear that the relative contributions of the three channels is different
for  $ J/\psi $ and $\Upsilon$ resonances. In case of $ J/\psi$ the dominant contributions
come from $\Gamma_{11}$ and $\Gamma_{22}$ corresponding to $HZZ$  and $HZ\gamma$ couplings respectively. 
The subleading contributions come from the interference terms $\Gamma_{12}$ and $\Gamma_{23}$. The
contribution $\Gamma_{33}$ coming from $Hq\bar{q}$ coupling is negligibly small. The
major contribution from Yukawa sector will come from the interference term $\Gamma_{23}$.
Therefore while probing the anomalous Yukawa couplings one should not neglect the contribution of
the interference terms over $\Gamma_{33}$.

However, in case of $\Upsilon$ the situation is quite different. The leading contribution comes only from the $\Gamma_{11}$ term
whereas $\Gamma_{12}$ and $\Gamma_{13}$ provide the subleading contributions.
The contribution of $\Gamma_{33}$ is now larger than $\Gamma_{22}$ but still negligibly small
compared to $\Gamma_{11}$. Again as before while probing anomalous Yukawa coupling the 
effect of interference terms can not be neglected.

As discussed above the rare Higgs decays $H \to Z V$ are sensitive not only
to $HZZ$ coupling but also to $HZ\gamma$ and $Hq\bar{q}$ couplings. Moreover, 
depending on nature of $V$, the contribution of various Higgs couplings to the branching fractions 
vary significantly. These decay modes have potential to provide information 
complimentary to $H \to Z Z^{*} \to 4 \ell$ ``golden channel'' and $HZ\gamma$. Furthermore, anomalous nature
of Yukawa couplings will be exhibited primarily through the interference terms.
Also, $H \to Z V$ decays followed by subsequent 
decays of $Z$ and $V$ into pair of leptons will provide a experimentally clean channel 
that can be used to probe them in future colliders or high luminosity LHC runs.
In next section we will discuss  the angular analysis technique which provide a powerful tool
for probing such couplings.


\section{Angular analysis and Observables for $H \to Z V \to 4 \ell$ decays}
\label{sec:anghvz}

In this section we formalize the necessary technique to probe $HZV$ vertex.
We start with writing down the general structure of the vertex and the different helicity amplitudes for
$H \to Z V$ decays. Combining all the channels, the general Lorentz and gauge invariant structure of the $HZV$ vertex can be written as
\begin{equation}\label{eq:VHZZ0}
   V^{\alpha \beta}_{HZV}=
  \bigg( a_1 \, g^{\alpha \beta} + a_2 \, P^{\alpha}P^{\beta} 
  + i a_3 \,\epsilon^{\alpha \beta \mu \nu} \; q_{1\mu}\,q_{2\nu} \bigg), 
\end{equation}
where $a_1$, $a_2$ and $a_3$ are vertex factors defined as
\begin{align}
a_1 & = -\left(\mathcal{K}_1\,a^{ZZ}_1+\mathcal{K}_2\,a^{Z\gamma}_1 q_1.q_2+\mathcal{K}_3\,a^{q\bar{q}}_1 q_1.q_2\right),\label{a1} \\
a_2 & = \left(\mathcal{K}_1~a^{ZZ}_2+\mathcal{K}_2~a^{Z\gamma}_1+\mathcal{K}_3 a^{q\bar{q}}_1\right), \label{a2} \\
a_3 & = -\left(\mathcal{K}_1~a^{ZZ}_3+\mathcal{K}_2~a^{Z\gamma}_3+\mathcal{K}_3 a^{q\bar{q}}_3\right). \label{a3}
\end{align}
and $P$, $q_1$, $q_2$ are the four momenta of the Higgs boson, $Z$ and $V$ respectively.
The couplings $a_1$, $a_2$ and $a_3$ can be extracted via angular asymmetries discussed below.
Any deviation from SM values will indicate anomalous nature of $H \to Z V$ decay.
 
 The decay under consideration can be expressed in terms of three helicity 
amplitudes $\mathcal{A}_{L}$, $\mathcal{A}_{\parallel}$ and $\mathcal{A}_{\perp}$ defined
in the transversity basis as
\begin{align}
  \mathcal{A}_{L} &= (M_H^2 - M_Z^2 - M_V^2)\, a_1 + M_H^2\, X^2 \, a_2, \label{eq:AL}\\
  \mathcal{A}_{\parallel} &= \sqrt{2 }M_Z M_V \, a_1, \label{eq:AA}\\
  \mathcal{A}_{\perp} &= \sqrt{2}M_Z M_V~M_H\, X  \, a_3, \label{eq:AP}
\end{align}
where $M_H$, $M_Z$ and $M_V$ are masses of $H$, $Z$ and $V$ respectively with 
\begin{eqnarray}\label{eq:X}
  X&=&\dsp\frac{\sqrt{\dsp\lambda(M_H^2,M_Z^2,M_V^2)}}{\dsp 2M_H}
\end{eqnarray}
where 
$\lambda(x,y,z)= x^2+y^2+z^2-2\,x\,y-2\,x\,z-2\,y\,z$.
The helicity amplitudes $\mathcal{A}_{L}$, $\mathcal{A}_{\parallel}$ and $\mathcal{A}_{\perp}$ have 
definite parity properties. $\mathcal{A}_{L}$, $\mathcal{A}_{\parallel}$ are $CP$ even in nature 
where as $\mathcal{A}_{\perp}$ is $CP$ odd.

The full angular distribution for $H \to Z_{(\ell^+ \ell^-)} V_{(\ell^+ \ell^-)}$ is given 
by following expression

\begin{figure}[htb!]
\centering
\includegraphics[width=.9 \textwidth]{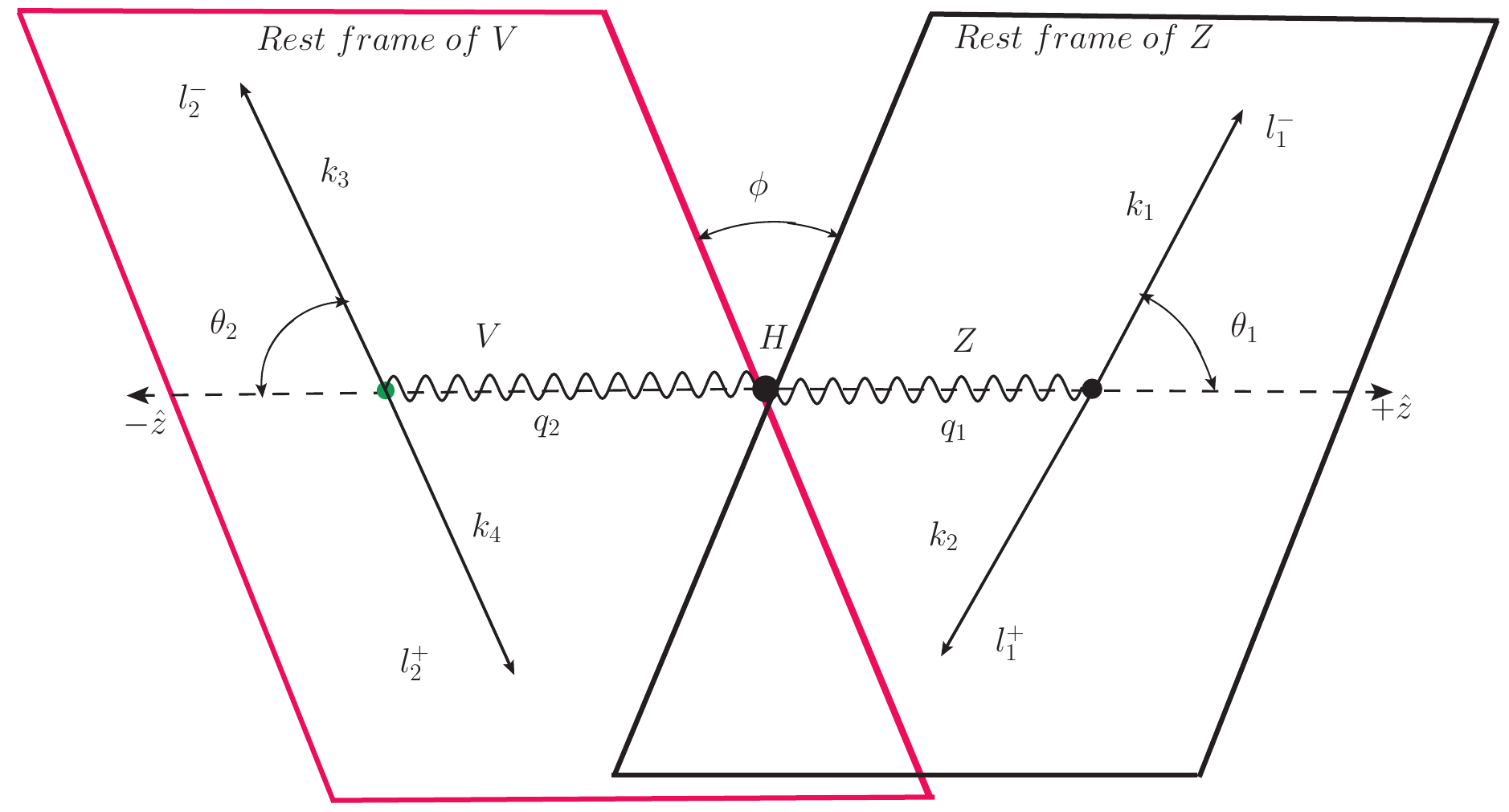}
\caption{The definition of the polar angles ($\theta_1$ and $\theta_2$)
  and the azimuthal angle ($\phi$) in the decay $H\to Z + V \to
  (\ell_1^- + \ell_1^+) + (\ell_2^- + \ell_2^+),$ where $\ell_1,
  \ell_2 \in \{ e,\mu \}$ and three momentum
  $\vec{k}_1 = -\vec{k}_2$ and $\vec{k}_3 = -\vec{k}_4$. The lepton pair
  $\ell_1^\pm$ goes back to back in the rest frame of $Z$, whereas  
  lepton pair $\ell_2^\pm$ goes back to back in the rest frame of $V$.}
\label{fig:refframe}
\end{figure}

\begin{widetext}
\begin{align}\label{angdist}
 & \frac{8\pi}{\Gamma}\frac{d^3\Gamma}{ d\cos{\theta_1} \;
    d\cos{\theta_2} \; d\phi} = 1 +\frac{|f_{\parallel}|^2-|f_{\perp}|^2}{4}
  \cos\,2\phi\big(1-P_2(\cos\theta_1)\big)
  \big(1-P_2(\cos\theta_2)\big)
\nn \\
&+\frac{1}{2} \Im(f_{\parallel} f_{\perp}^*)\sin\,2\phi \big(1-P_2(\cos\theta_1)\big) \big(1-P_2(\cos\theta_2)\big)
+\frac{1}{2}(1-3
  \modulus{f_L}^2)\,\big(P_2(\cos\theta_1)+P_2(\cos\theta_2)\big)\nn\\
&+\frac{1}{4}(1+3\modulus{f_L}^2)\,P_2(\cos\theta_1)P_2(\cos\theta_2)
  +\frac{9}{8\sqrt{2}} \left( \Re(f_L
    f_{\parallel}^*)\,\cos\phi + \Im(f_L f_{\perp}^*)\,\sin\phi
  \right)
  \sin2 \theta_1 \sin 2\theta_2 \nn\\
&-\eta\frac{9}{2\sqrt{2}}\Re(f_L f_{\perp}^*)
  \cos\theta_2\cos\phi \sin\theta_1\sin\theta_2
+\eta\frac{9}{2\sqrt{2}}\Im(f_L f_{\parallel}^*)
  \cos\theta_2\sin\phi
  \sin\theta_1\sin\theta_2\nn\\
&+\eta\frac{3}{2} \Re(f_{\parallel}
  f_{\perp}^*)\big(\cos\theta_2 (2 + P_2(\cos\theta_1))
  - \cos\theta_1
  (2 + P_2(\cos\theta_2))\big) 
\end{align}
\end{widetext}
where the angle $\theta_1$($\theta_2$) is the angle between three momenta of $\ell^+$ in $Z$($V$) rest frame 
and the direction of three momenta of $Z$($V$) in $H$ rest frame as shown in Fig.~\ref{fig:refframe}. 
The angle $\phi$ is defined as the angle between 
the normals to the planes defined by  $Z\to \ell^+ \ell^-$ and  $V\to \ell^+ \ell^-$ in $H$ rest frame.
The expressions for {\it{helicity fractions}} $f_L$, $f_{\parallel}$ and $f_{\perp}$ are given in the appendix.

 Integrating Eq.\eqref{angdist} with respect to the angles
$\cos{\theta_1}$ or $\cos{\theta_2}$ or $\phi$, one can obtain following
uniangular distributions:
\begin{align}
  \frac{1}{\Gamma}\frac{d\Gamma}{d\cos\theta_1} &=
  \frac{1}{2} + t_2\,P_2(\cos\theta_1) - t_1
  \cos\theta_1,  \label{eq:t1dist} \\
  \frac{1}{\Gamma}\frac{d\Gamma}{ d\cos\theta_2} &=
  \frac{1}{2} + t_2\,P_2(\cos\theta_2) , \label{eq:t2dist} \\
  \frac{2\pi}{\Gamma}\frac{d\Gamma}{d\phi} &= 1 +
  u_2\,\cos\,2\phi + v_2\,\sin\,2\phi  \label{eq:phidist}
\end{align}
where  $P_2(\cos\theta_{1,2})$ are second degree Legendre Polynomial and
\begin{align}
\label{eq:t1}
t_1 &=\frac{3}{2} \, \eta \mathrm{\Re}(f_{\parallel} f_{\perp}^*),\\
\label{eq:t2}
t_2 &=\frac{1}{4} (1-3\modulus{f_L}^2),\\
\label{eq:v2}
v_2 &=\frac{1}{2}\,\mathrm{Im}(f_{\parallel}f_{\perp}^*),\\
\label{eq:u2}
u_2 &=\frac{1}{4} (|f_{\parallel}|^2-|f_{\perp}|^2).
\end{align}
The uniangular distributions in Eq.\eqref{eq:t1dist}, Eq.\eqref{eq:t2dist} and Eq.\eqref{eq:phidist} 
will provide unique probe to study the $H\to Z V$ couplings. 
The observables $t_1$, $t_2$, $u_2$ and $v_2$ can be extracted using following 
asymmetries:

\begin{align}
&  t_1 =  \frac{1}{\Gamma} \left( \int_{-1}^{0} - \int_{0}^{+1} \right)
   \frac{d\Gamma}{ d\cos\theta_1} \, d\cos\theta_1, 
\label{eq:T10asym}\\%
&  t_2 = \!\frac{4}{3 \Gamma}\! \left( \int_{-1}^{-\frac{1}{2}} -
  \int_{-\frac{1}{2}}^{+\frac{1}{2}} + \int_{+\frac{1}{2}}^{+1} \right)
\frac{d\Gamma}{d\cos\theta_{1,2}}  d\cos\theta_{1,2}
\label{eq:T20asym}\\%
&  v_2 = \frac{\pi}{2\Gamma} \left( \int_{-\pi}^{-\frac{\pi}{2}} -
  \int_{-\frac{\pi}{2}}^{0} + \int_{0}^{+\frac{\pi}{2}} -
  \int_{+\frac{\pi}{2}}^{+\pi} \right) \frac{d\Gamma}{d\phi}d\phi,
   \label{eq:V20asym}\\%
&  u_2 = \frac{\pi}{2 \Gamma} \left( \int_{-\pi}^{-\frac{3\pi}{4}} -
  \int_{-\frac{3\pi}{4}}^{-\frac{\pi}{4}} +
  \int_{-\frac{\pi}{4}}^{\frac{\pi}{4}}
  -\int_{\frac{\pi}{4}}^{\frac{3\pi}{4}} + \int_{\frac{3\pi}{4}}^{\pi}
  \right)\frac{d\Gamma}{d\phi} d\phi.
\label{eq:U20asym}
\end{align}

The observables $t_1$, $t_2$, $u_2$, $v_2$ are functions of $a_1$, $a_2$, 
$a_3$ and hence their measurements will allow us to probe $H\to Z V$ coupling.
In SM $t_1$, $t_2$, $u_2$, $v_2$ have unique values which can be computed using 
the SM values of the couplings $a_1$, $a_2$, $a_3$ given in Eq.\eqref{a1}, 
Eq.\eqref{a2} and Eq.\eqref{a3}. The anomalous nature, if any, of $a_1$, $a_2$, $a_3$ will 
show up in the observables as deviation from their SM values.

As discussed in Section \ref{sec:direct}, the rare Higgs decays are sensitive to $HZZ$, $HZ\gamma$
and $Hq\bar{q}$ couplings. Therefore, any deviation of the observables $t_1$, $t_2$, $u_2$, $v_2$ from their SM values
can not a priori be attributed to anomalous nature of any one sector. However, when taken in conjugation with 
other decays like $H\to Z Z^{*}\to 4\ell$ they can provide complimentary information about $HZ\gamma$
and $Hq\bar{q}$ couplings. For example, if any hint of anomalous nature is observed in $H\to Z Z^{*}\to 
4\ell$ decay, one expects to see corresponding deviations in the observables of $H \to ZV$ for both $J/\psi$ and $\Upsilon$.
On the other hand if $H\to Z Z^{*}\to 4\ell$ decay observables turn out to be consistent with the SM 
values then $HZZ$ contribution in rare decays should also be SM like. In such a scenario any observed anomaly in $H \to Z V$
can only arise from either $HZ\gamma$ or $H q\bar{q}$ couplings. As the magnitude of their contributions in $H \to Z\, J/\psi$ 
and $H \to Z\, \Upsilon$  are quite different, this fact can be exploited to further narrow down the origin of the anomalous behaviour. 
Moreover, for Higgs decay to both $J/\psi$ and $\Upsilon$, the effect of any anomaly in Yukawa sector will predominantly modify the interference terms.

In principle Higgs can have anomalous couplings in more than one sector. If so, it will be relatively more difficult 
to make any definite conclusions about the relative contributions of the three sectors to the anomalous couplings of Higgs in
$H \to Z V$ decays. 

In several NP scenarios parity violating anomalous couplings can arise.
Depending on the NP scenario under consideration, they can arise either only in
Yukawa sector or only in gauge sector or in both sectors simultaneously.
To elaborate this we consider three benchmark scenarios which are tabulated in Table~\ref{benchmark} and find the corresponding
uniangular distributions. The Benchmark-I scenario is for the SM i.e. $a^{ZZ}_1=a^{Z\gamma}_1=a^{q \bar{q}}_1=1$
and $a^{ZZ}_2=a^{ZZ}_3=a^{Z\gamma}_3=a^{q \bar{q}}_3=0$. Benchmark-II and Benchmark-III scenario are 
characterized by the non zero parity violating terms $a^{Z\gamma}_3$ and  $a^{q \bar{q}}_3$  along with 
$a^{ZZ}_1=a^{Z\gamma}_1=a^{q \bar{q}}_1=1$ respectively.
\begin{table}[hbtp!]
\centering
\begin{tabular}{|c|c|c|c|}
\hline
Couplings                          & Benchmark-I                &   Benchmark-II         &  Benchmark-III      \\
\hline
\hline
 $a^{ZZ}_1$                       &    1                      &        1             &        1               \\
 $a^{Z\gamma}_1$                  &    1                      &        1             &        1               \\
 $a^{q \bar{q}}_1$                &    1                      &        1             &        1               \\
 $a^{ZZ}_2$                       &    0                      &        0             &        0               \\
 $a^{ZZ}_3$                       &    0                      &        0             &        0                \\
 $a^{Z\gamma}_3$                  &    0                      &        2+ 2 i        &        0                \\
 $a^{q \bar{q}}_3$                &    0                      &        0             &        5 + 5 i                \\
\hline
\end{tabular}
\caption{Three Benchmark scenarios with Benchmark-I conforms SM. In Benchmark-II we have allowed non zero
value for parity violating term $a^{Z\gamma}_3$ however for Benchmark-III the the parity violating term $a^{q \bar{q}}_3$
is kept non zero.}
\label{benchmark}
\end{table}

\begin{figure*}[hbtp!]
\centering
\subfloat[]{\includegraphics[width=.48\textwidth]{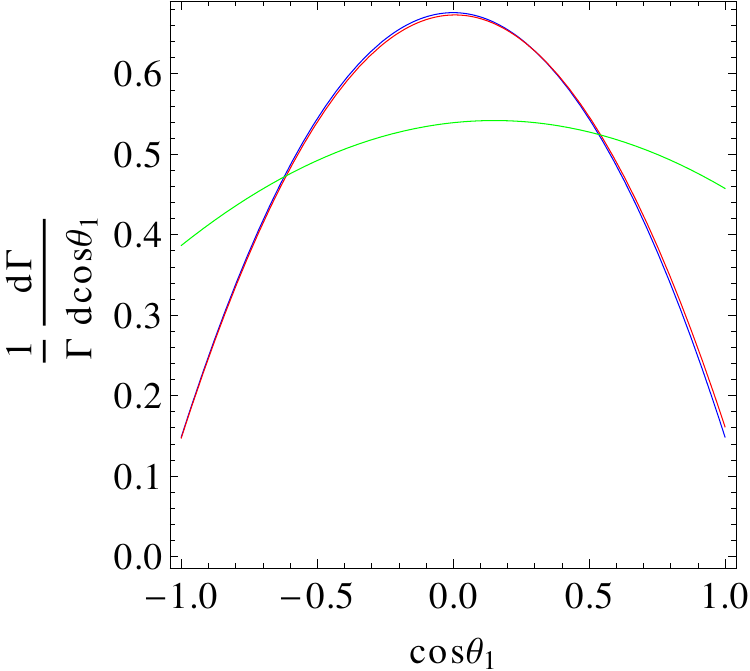}}
\subfloat[]{\includegraphics[width=.48\textwidth]{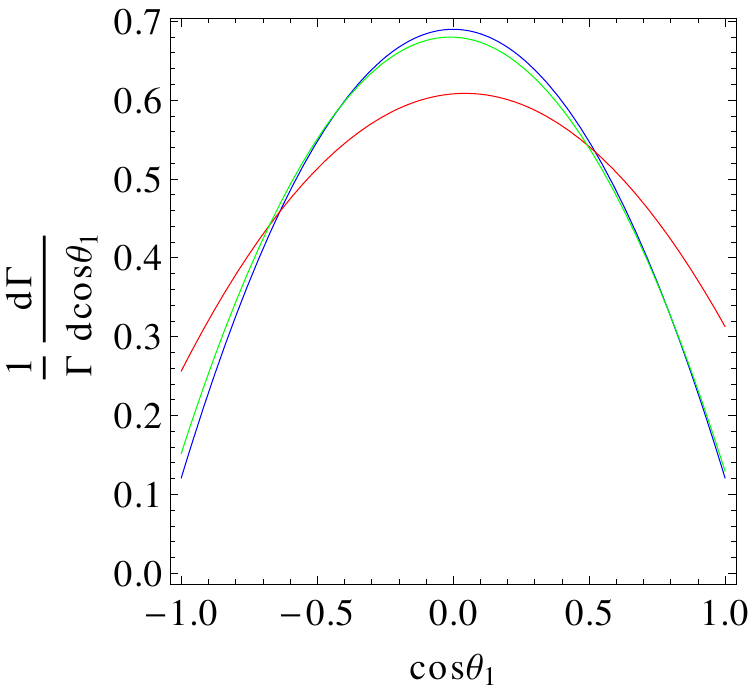}} 
\caption{The angular distribution $\frac{1}{\Gamma}\frac{d\Gamma}{d\cos\theta_1}$ vs $\cos\theta_1$ for 
   $J/\psi$(a) and $\Upsilon$(b). The blue line corresponds to the Benchmark-I(SM) scenario where as green and red lines correspond 
    to the Benchmark-II and the Benchmark-III scenarios.} 
\label{fig:fig1} 
\end{figure*}

\begin{figure*}[hbtp!]
\centering
\subfloat[]{\includegraphics[width=.48\textwidth]{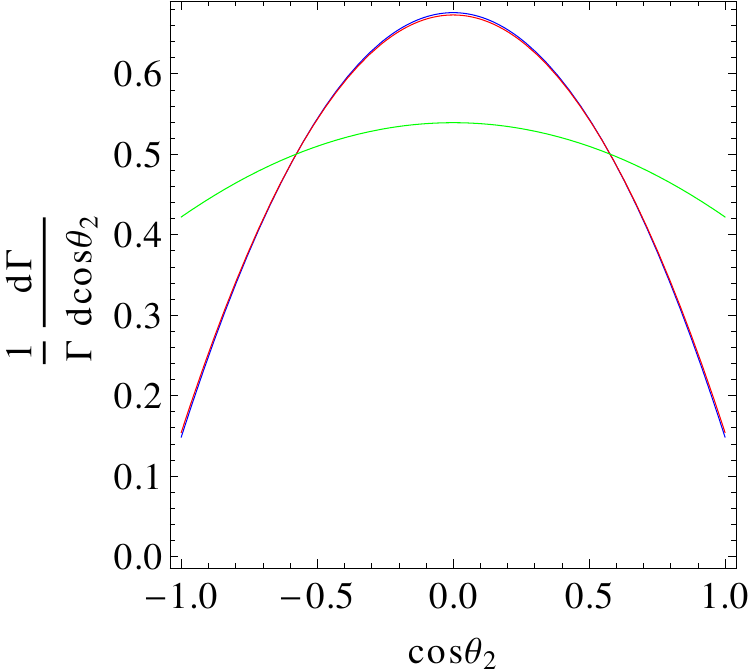}} 
\subfloat[]{\includegraphics[width=.48\textwidth]{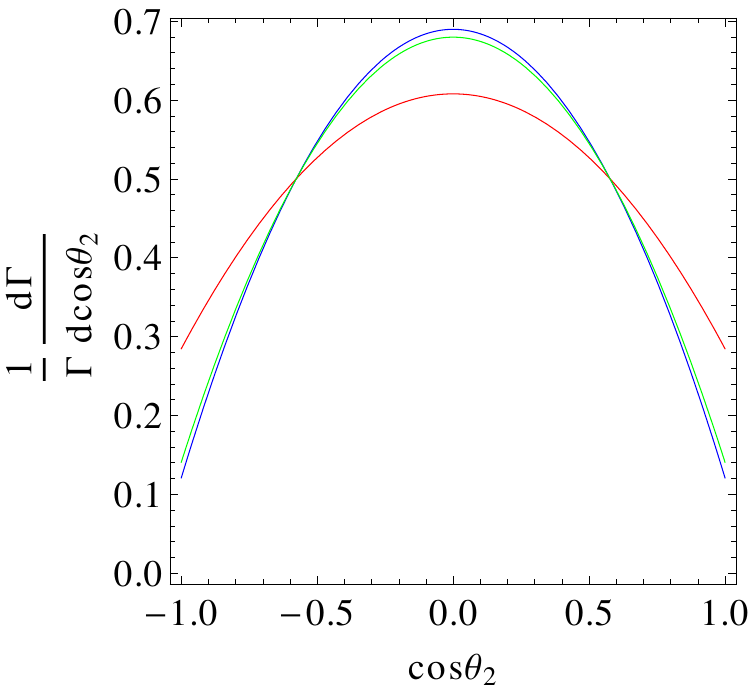}}
\caption{The angular distribution $\frac{1}{\Gamma}\frac{d\Gamma}{d\cos\theta_2}$ vs $\cos\theta_2$ for 
   $J/\psi$(a) and $\Upsilon$(b). The blue line corresponds to the Benchmark-I(SM) scenario where as green and red  lines correspond 
    to the Benchmark-II and the Benchmark-III scenarios.} 
\label{fig:fig2} 
\end{figure*}

\begin{figure*}[hbtp!]
\centering
\subfloat[]{\includegraphics[width=.48\textwidth]{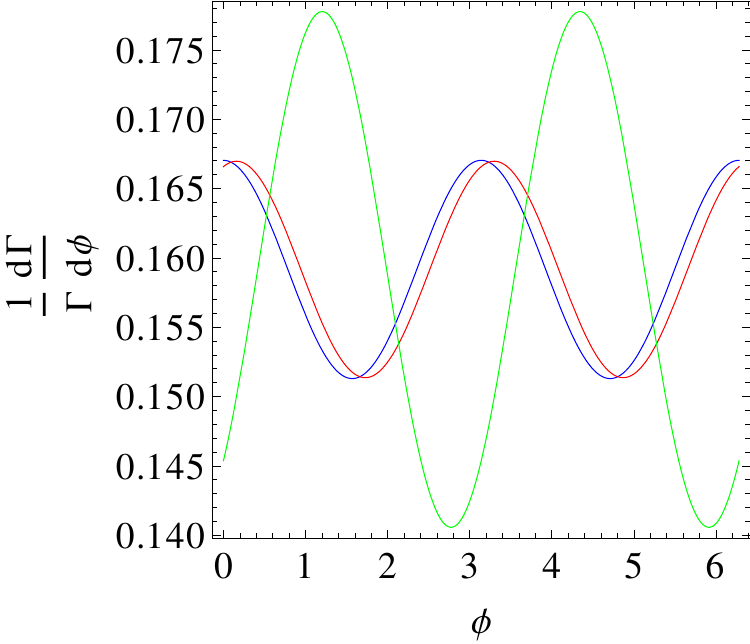}} 
\subfloat[]{\includegraphics[width=.48\textwidth]{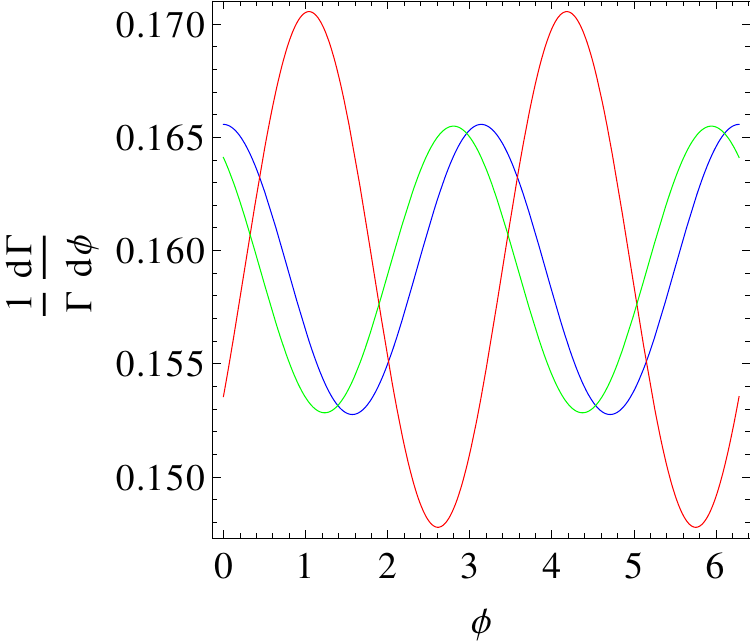}}
\caption{The angular distribution $\frac{1}{\Gamma}\frac{d\Gamma}{d\phi}$ vs $\phi$ for 
    $J/\psi$(a) and $\Upsilon$(b). The blue line corresponds to the Benchmark-I(SM) scenario where as green and red lines correspond 
     to the Benchmark-II and the Benchmark-III scenarios.} 
\label{fig:fig3} 
\end{figure*}

 The effect of  benchmark scenarios on uniangular distributions  $\frac{1}{\Gamma}\frac{d\Gamma}{d\cos\theta_1}$, 
$\frac{1}{\Gamma}\frac{d\Gamma}{d\cos\theta_2}$ and $\frac{1}{\Gamma}\frac{d\Gamma}{d\phi}$ vs $\phi$ are shown in 
Fig.~\ref{fig:fig1}, Fig.~\ref{fig:fig2} and Fig.~\ref{fig:fig3} respectively.

If the parity violating NP term in $H \to ZV$ decay originates from $HZ\gamma$ vertex
 the angular distributions for $H \to Z J/\psi$ will deviate from SM distributions of Benchmark-I.This are
shown in Fig.~\ref{fig:fig1}(a), Fig.~\ref{fig:fig2}(a) and Fig.~\ref{fig:fig3}(a) where the Benchmark-II green line
deviates from the blue SM line. However if parity violating term for $H \to Z J/\psi$ generates from $a^{q \bar{q}}_3$ term
(i.e. for Benchmark-III), we will not see any significant deviations in any of the angular distributions. 
This is because the second highest contribution to the total decay width $\Gamma$ (and hence also in the angular distributions)
comes from $HZ\gamma$ diagram which is given by $\mathcal{B}r_{\Gamma_{22}}$ in Table~\ref{tab}.
As the term $a^{Z\gamma}_3$ is normalized to the term $\mathcal{K}_2$ in Eq.\eqref{a3} the effect is repeated even when
the non zero parity violating term $a^{Z\gamma}_3$ is present.
For the $H \to Z \Upsilon$ decays, the Benchmark-III (red line) deviates more from SM(blue line) than the Benchmark-II (green line).
However for most general case NP can arise from any of these sectors and to completely disentangle the origin of such contributions
depend on the precise measurement of the $HZZ$ vertex and $HZ\gamma$ vertex.
  
\section{Conclusion}\label{sec:conclusion}

 In this work we have looked at the potential of probing anomalous Higgs 
couplings via the rare Higgs decays $H \to Z V$; $V = J/\psi, \Upsilon$.
The rare Higgs decays provide an unique opportunity to probe anomalous 
Higgs couplings in both gauge and Yukawa sectors. 
In this work we have computed the relative strength of 
different sectors contributing to the branching fraction 
of the $H \to ZV$ decays in SM. The relative contribution of different diagrams 
and their interference terms varies
significantly depending on whether $V$ is $J/\psi$ or $\Upsilon$. We find
that in SM the branching fraction of $H \to Z V$ decays are small.
Hence they provide a very sensitive tool to probe physics beyond 
SM for the scenarios where the branching fraction of $H \to Z V$ 
decays is enhanced by the new physics contributions e.g. via anomalous Yukawa
couplings to quarks.
 
The subsequent decay of both $Z$ and $V$ into pair of leptons make 
$H\to Z V$ decays experimentally clean for collider studies. Furthermore,
one can fully reconstruct the phase space of Higgs from its four 
lepton final state and can find several kinematic variables to study the $HZV$ vertex.
In particular one can construct three angles from the four lepton final 
state and use them as kinematic variables to extract out anomalous Higgs couplings
in $H\to Z V$ decays. 

The $H\to Z V$ decays receive contributions from gauge as well as 
Yukawa sectors and depending on the nature of $V$, can be sensitive 
to anomalous couplings in more than one sector. If $HZV$ couplings are 
found to be anomalous, contrary to several previous claims, one cannot 
immediately conclude that they necessarily imply anomalous Yukawa couplings. However when 
combined with other decay modes of Higgs such as $H \to ZZ^* \to 4 \ell$, $H \to Z \gamma \to \ell_1^+ \ell_1^- \gamma$ etc.,
the rare Higgs decays can help us to unravel the origin of anomalous couplings in either the gauge 
or Yukawa sectors. We finally conclude that the rare  
Higgs decays and angular asymmetries will play an essential role in 
probing potential New Physics contributions, in high luminosity LHC runs
as well as in future colliders.

\acknowledgments
We thank Arjun Menon, Dibyakrupa Sahoo and Rahul Sinha for many fruitful discussions.

\appendix
\section{}

{\it{Helicity fractions}} $f_L$, $f_{\parallel}$ and
$f_{\perp}$ are defined as
\begin{equation}
  \dsp f_\lambda=\frac{\mathcal{A}_\lambda}{\sqrt{ \modulus{\mathcal{A}_{L}}^2 +
      \modulus{\mathcal{A}_{\parallel}}^2 + \modulus{\mathcal{A}_{\perp}}^2}}, 
\end{equation}
where $\lambda \in \{ L, \parallel, \perp \}$ 
and
\begin{align}
  \Gamma = \mathcal{N} \left(
    \modulus{\mathcal{A}_{L}}^2 + \modulus{\mathcal{A}_{\parallel}}^2 +
    \modulus{\mathcal{A}_{\perp}}^2 \right) \label{eq:gamma},%
\end{align}
with 
\begin{align}
  \mathcal{N} &= \frac{1}{2^{7}} \; \frac{1}{9\pi^3} \;
  \frac{X}{M_H^2} \; \frac{\mathcal{K}^2}{\Gamma_Z M_Z \Gamma_V M_V} {a_v}^2
  \; (v_{\sss\ell}^2+a_{\sss\ell}^2)%
\end{align}
$v_{\sss\ell}=2 I_{3\ell}-4e_{\sss\ell} \sin^2\theta_W$, $a_{\sss\ell}=2 I_{3\ell}$,
$a_v=\frac{4 \pi Q^f \alpha f_V}{\sqrt{3} M_V}$
and $\eta$ is defined as $\eta=\frac{2 v_{\sss\ell} a_{\sss\ell}}{v_{\sss\ell}^2+a_{\sss\ell}^2}$.

\end{document}